\documentstyle[aps,prl,twocolumn,floats,epsfig]{revtex}
\begin{document}

\bibliographystyle{prsty}

\title{ \begin{flushleft}
{\small
\sc 
Volume 79, Number 22
\hfill
PHYSICAL REVIEW LETTERS 
\hfill 
1 December 1997, 4469--72} \\
\vspace{-2mm}
\_\hrulefill \\
\end{flushleft}  
First- and Second-Order Transitions between Quantum and Classical Regimes \\
 for the Escape Rate of a Spin System
\vspace{-1mm}
}

\author{E. M. Chudnovsky 
\renewcommand{\thefootnote}{\fnsymbol{footnote}}
\footnotemark[1]
and D. A. Garanin
\renewcommand{\thefootnote}{\fnsymbol{footnote}}
\footnotemark[2]
}

\address{
Department of Physics and Astronomy, City University of New York -- 
Lehman College,\\
 Bedford Park Boulevard West, Bronx, New York 10468-1589 \\
\smallskip
{\rm(Received 7 July 1997)}
\bigskip\\
\parbox{14.2cm}
{\rm
We have found a novel feature of the bistable large-spin model described by
the Hamiltonian ${\cal H} = -DS_z^2 - H_xS_x $.
The crossover from thermal to quantum regime for the escape rate 
can be either first ($H_x<SD/2$)
or second ($SD/2<H_x<2SD$) order, that is, sharp or smooth,
depending on the strength of the transverse field.
This prediction can be tested experimentally in molecular
magnets like Mn$_{12}$Ac.
\smallskip
\begin{flushleft}
\end{flushleft}
} 
} 
\maketitle

Transitions between two states in a bistable system can occur
either due to the classical thermal activation or via quantum tunneling.
A rigorous study of that problem was begun by Kramers
\cite{kra40} and WKB \cite{wen26,kra26,bri26}, 
and a review of the progress that
followed can be found in Ref. \cite{haetalbor90}. 
At high temperature the transition rate follows the Arrhenius law,
$\Gamma \sim \exp(-\Delta U/T)$, with $\Delta U$ being the hight
of the energy barrier between the two states.
In the limit of $T \to 0$, the transitions are purely quantum, 
$\Gamma \sim \exp(-B)$, with $B$ independent on temperature.
Due to the exponential dependence of the thermal rate on $T$,
the temperature $T_0$ of the crossover from quantum to thermal regime
can be estimated as $T_0^{(0)} = \Delta U/B$.
For a quasiclassical particle in a potential $U(x)$, Goldanskii 
\cite{gol59} noticed the possibility of a more accurate definition,
$T_0^{(2)} = \hbar/\tau_0$, where $\tau_0$ is the period of
small oscillations near the bottom of the inverted potential, $-U(x)$.
Below $T_0^{(2)}$, thermally assisted tunneling occurs from the
excited levels, that reduces to the tunneling from the
ground-state level at $T=0$.
Above $T_0^{(2)}$ quantum effects are small and the transitions
occur due to the thermal activation to the top of the barrier.
Affleck \cite{aff81} demonstrated that the two regimes regimes
smoothly join at $T=T_0^{(2)}$.
Larkin and Ovchinnikov \cite{larovc83} called this situation the
second-order phase transition from classical to quantum behavior.
This means that for $\Gamma$ written as 
$\Gamma \sim \exp(-\Delta U/T_{\rm eff})$, the dependence of both
$T_{\rm eff}$ and its first derivative on $T$ are continuous at    
$T=T_0^{(2)}$.
This situation is not generic, however.
The transition between the two regimes can also be of the first order
\cite{larovc83,chu92}, i.e., more abrupt, with $dT_{\rm eff}/dT$
discontinuous at a certain temperature $T_0^{(1)} > T_0^{(2)}$.
Chudnovsky derived the criterium allowing one to establish
whether first- or second-order transition takes place, based on
the shape of the potential $U(x)$.
Commonly studied potentials $U = -x^2 + x^4$ and $U = -x^2 +
x^3$ yield the second-order transition.
Physically relevant potentials which would exhibit the
first-order transitions were not known.
In this work we show that spin systems readily accessible in the
experiment possess both first- and second-order transitions
between the classical and quantum behavior of the escape rate.
The order of the transition in these systems can be controlled
by external magnetic field.

Consider a spin system described by the Hamiltonian
\begin{equation}\label{sham}
{\cal H} = -DS_z^2 - H_xS_x 
\end{equation}
where $S\gg 1$.
This model is generic for problems of spin tunneling studied by
different methods 
\cite{enzsch86,lvhsuto86,schwrelvh87,chugun88,zas90plazas90prb,gar91jpa}.
It is believed to be a good approximation for the molecular
magnet Mn$_{12}$Ac of spin $S=10$, intensively studied in last years
(see, e.g., Ref. \cite{sch97}).
In the quasiclassical approximation the transition rate is given
by
\begin{equation}\label{gamint}
\Gamma \sim \int \!\!dE\; W(E) e^{-(E-E_{\rm min})/T},
\end{equation}
where $W(E)$ is the probability of tunneling at an energy $E$ and $E_{\rm min}$
corresponds to the bottom of the potential.
This probability is defined via the imaginary-time action
\begin{equation}\label{we}
\qquad W(E) \sim e^{-S(E)}.
\end{equation}
With the accuracy to the exponent,
\begin{equation}\label{gammin}
\Gamma \sim e^{-F_{\rm min}/T},
\end{equation}
where $F_{\rm min}$ is the minimum of the effective ``free energy''
\begin{equation}\label{feff}
F \equiv E + TS(E) - E_{\rm min}
\end{equation}
with respect to $E$.

In order to obtain $S(E)$ for the Hamiltonian (\ref{sham}) we
will use the method of mapping the spin problem onto a particle
problem \cite{schwrelvh87,zas90plazas90prb,scharf74,zasulytsu83}.
The equivalent particle Hamiltonian is
\begin{equation}\label{schroed}
{\cal H} = -\frac{\nabla^2}{2m} + U(x), 
\end{equation}
where
\begin{equation}\label{ux}
U(x) = \left(S+\frac12\right)^2D(h_x^2 \sinh^2 x - 2h_x \cosh x),  
\end{equation}
and
\begin{equation}\label{mhx}
m \equiv \frac{1}{2D}, \qquad  h_x\equiv \frac{H_x}{(2S+1)D}.
\end{equation}
In the future we shall neglect 1/2 in comparison to $S\gg 1$.

The imaginary-time action is then given by the WKB expression
\begin{equation}\label{action}
S(E) = 2(2m)^{1/2}\int\limits_{-x(E)}^{x(E)} dx\; \sqrt{U(x)-E},
\end{equation}
where $\pm x(E)$ are the turning points for the particle
oscillating inside the inverted potential $-U(x)$.
The period of these oscilations, $\tau_p(E)=-dS(E)/dE$, depends
on energy.
Minimization of (\ref{feff}) gives
\begin{equation}\label{extrcond}
\tau_p(E) = \frac1T,
\end{equation}
the condition familiar from the quantum statistics \cite{fey72,aff81,larovc83}.
It determines the instanton trajectory that dominates the
transition rate at a temperature $T$.

The dependence of $\tau_p(E)$ on $E$ determines the kind of the
crossover from quantum tunneling to thermal activation \cite{chu92}.
If $\tau_p$ monotonically increases with the amplitude of
oscillations, i.e., with decreasing energy $E$, the transition
is of the second order.
This kind of the crossover has been intensively studied,
including the case of tunneling with dissipation 
\cite{calleg83,grawei84,larovc84,zwe85,rishaefre85}.
If, however, the dependence of $\tau_p(E)$ is non-monotonic, the
first-order crossover takes place.
Let us demonstrate that both kinds of the crossover exist for
our spin model, depending on the strength of the transverse field.
Expanding (\ref{ux}) near $x=0$, one obtains
\begin{eqnarray}\label{uxsmall}
&&
U(x) \cong U(0) + S^2D\bigg[-h_x(1-h_x)x^2 
\nonumber\\
&&
\qquad
\left.
{}+\frac{h_x}{3}\left(h_x-\frac14\right)x^4 + O(x^6)\right] ,
\end{eqnarray}
where the sixth-order term is positive.
The second-order term in (\ref{uxsmall}) is negative for
$h_x<1$, which corresponds to the existence of the energy barrier
\begin{equation}\label{uxbarr}
U_{\rm max} - U_{\rm min} = S^2 D(1-h_x)^2.
\end{equation}
For $h_x>1/4$ the fourth-order term in (\ref{uxsmall}) is
positive, i.e., $U(x)$ is of the form $-x^2+x^4$.
The inverted potential $-U(x)$ is hence of the type $x^2-x^4$,
which results in the increase of $\tau_p$ with the oscillation
amplitude (i.e., with lowering the energy $E$) and to the second-order 
transition \cite{chu92}.
At $h_x<1/4$ the anharmonicity of $-U(x)$ has the
opposite sign, $-U(x) \sim  x^2+x^4$, which leads to the
decrease of $\tau_p$ when lowering $E$ for energies below the
top of the barrier.
However, with further lowering of $E$ the period $\tau_p$ begins to
increase and diverges logarithmically for $E$ approaching the
bottom of the potential.
This non-monotonic behavior of $\tau_p(E)$ leads to the
first-order transition from the thermally activated escape to the
quantum escape \cite{chu92}.

\begin{figure}[t]
\unitlength1cm
\begin{picture}(11,7)
%
\centerline{\epsfig{file=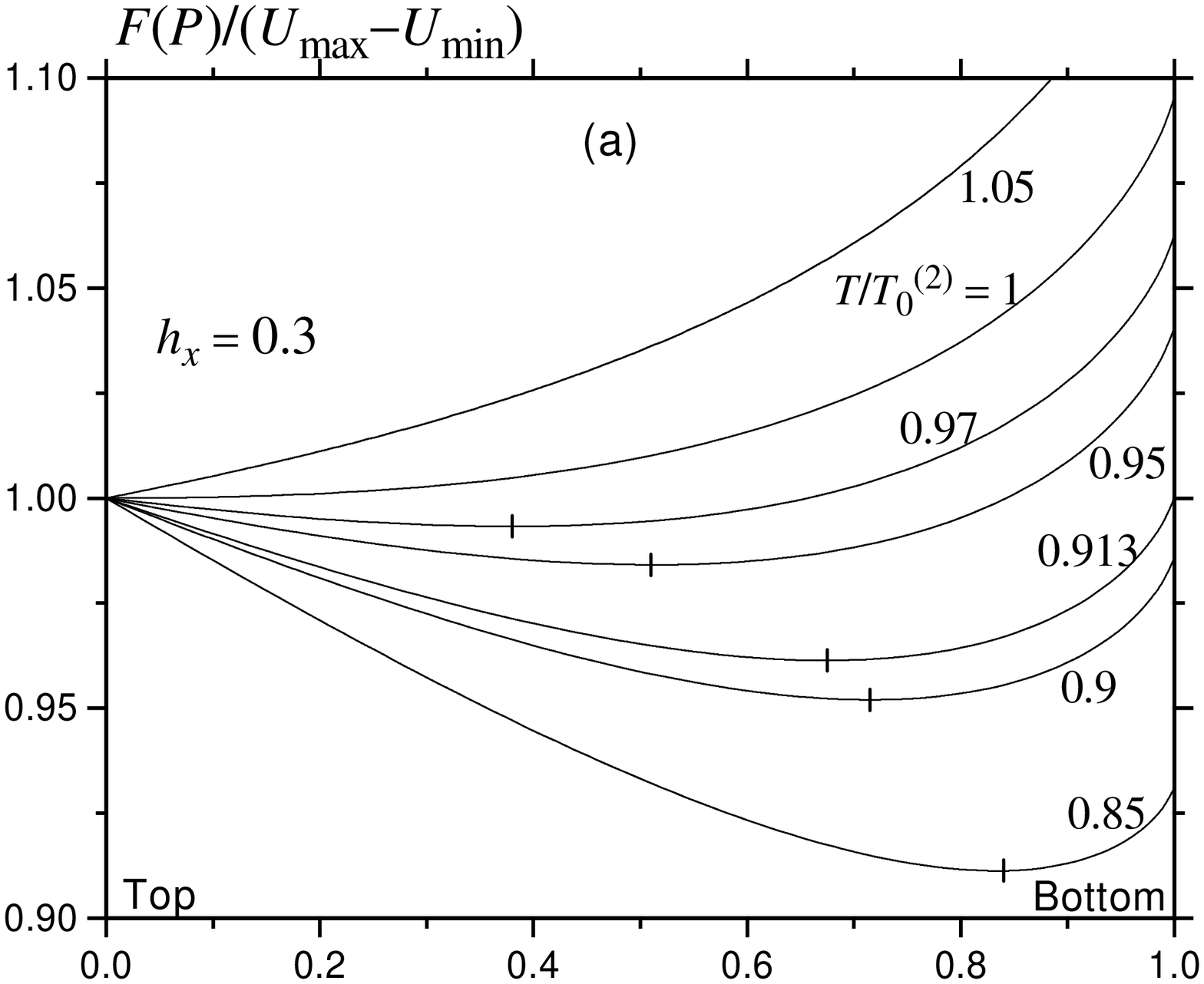,angle=-90,width=12cm}}
\end{picture}
\begin{picture}(11,6)
%
\centerline{\epsfig{file=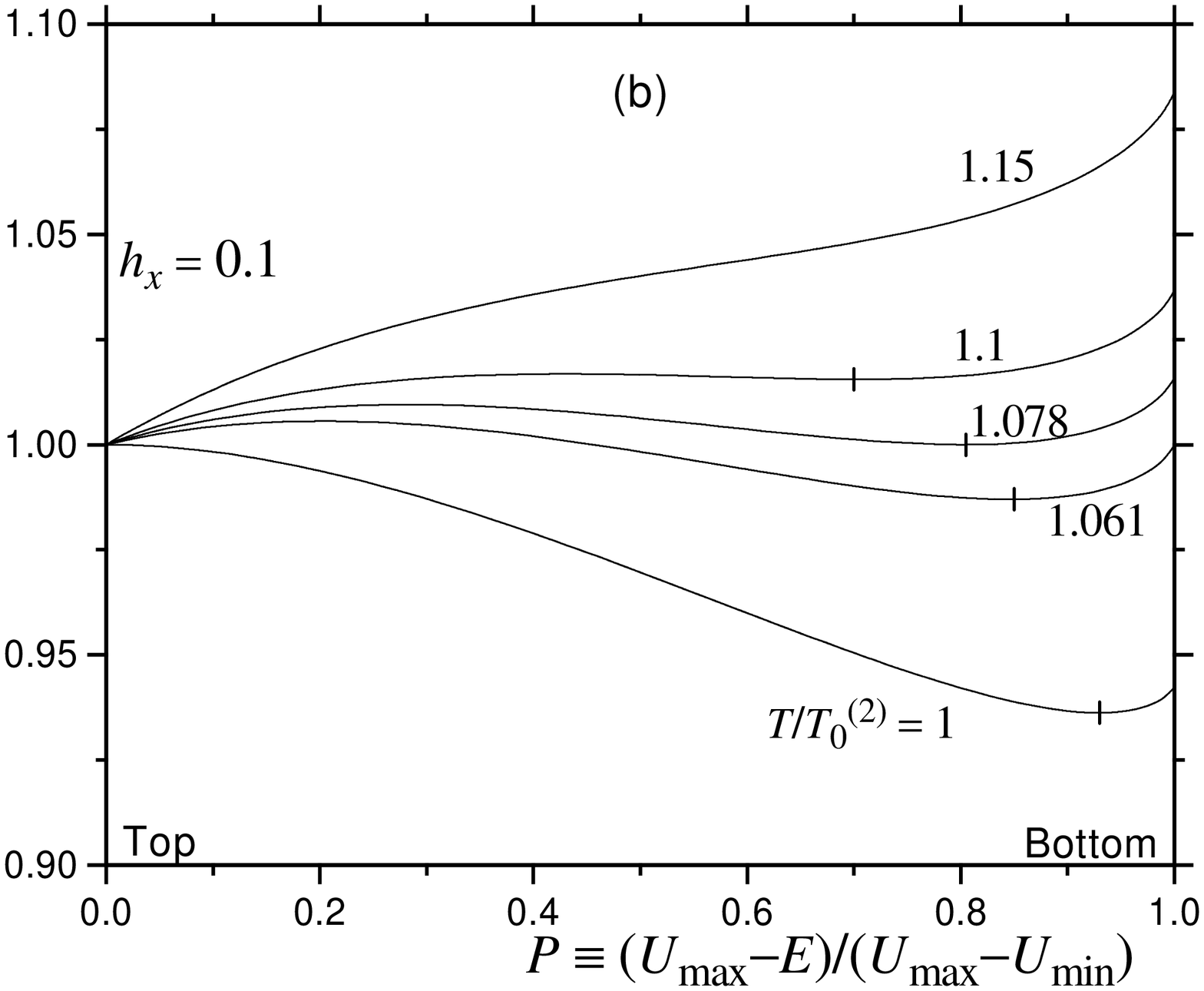,angle=-90,width=12cm}}
\end{picture}
\caption{ \label{tat_f1} 
Effective ``free energy'' for the escape rate:
(a) -- $h_x\equiv H_x/(2SD) = 0.3$, second-order transition;
(b) -- $h_x = 0.1$, first-order transition.
}
\end{figure}

In the case of the second-order transition the crossover occurs at
temperature
\begin{equation}\label{t02}
T_0^{(2)} = \frac{\tilde\omega_0}{2\pi} = \frac{SD}{\pi}\sqrt{h_x(1-h_x)} , 
\end{equation}
where $\tilde\omega_0=\sqrt{|U''(0)|/m}$ is the instanton frequency 
\cite{gol59,aff81}.
It is convenient to introduce dimensionless temperature and energy variables
\begin{equation}\label{dimlessvar}
\theta\equiv\frac{T}{T_0^{(2)}},
\qquad P\equiv \frac{ U_{\rm max} - E }{ U_{\rm max} -U_{\rm min} } .
\end{equation}
The effective free energy (\ref{feff}) near the top of the
barrier ($P<<1$) can be calculated with the use of Eqs. (\ref{action})
and (\ref{uxsmall}) and reads
\begin{eqnarray}\label{fpsmall}
&&
\frac{F(P)}{U_{\rm max} -U_{\rm min}} \cong 1 + (\theta-1)P 
+ \frac{\theta}{8} \left(1-\frac{1}{4h_x}\right)P^2
\nonumber\\
&&
\qquad
+ \frac{3\theta}{64} \left(1-\frac{1}{3h_x}+\frac{1}{16h_x^2}\right)P^3
+ O(P^4).
\end{eqnarray}
The analogy with the Landau theory of phase transitions,
described by $F = a\psi^2 + b\psi^4 + c\psi^6$, now becomes apparent.
The factor in front of $P$ (the Landau coefficient $a$) changes
the sign at the phase transition temperature $T=T_0^{(2)}$.
The factor in front of $P^2$ (the Landau coefficient $b$) changes the sign
at the field value $h_x=1/4$ determining the phase boundary
between the first- and the second-order transitions, as has been
already noticed from Eq. (\ref{uxsmall}).
The factor in front of $P^3$ (the Landau coefficient $c$) 
remains always positive.
The numerically computed dependence of $F$ on $P$ for the entire range of 
energy is plotted in Fig. \ref{tat_f1}.     
At $h_x=0.3$ (Fig. 1a) the minimum of $F$ remains $U_{\rm max} -U_{\rm min}$
for all $T>T_0^{(2)}$. 
Below $T_0^{(2)}$ it continuously shifts from
the top to the bottom of the potential as temperature is lowered.
This corresponds to the second-order transition from thermal
activation to thermally assisted tunneling.
At $h_x=0.1$ (Fig. 2b), however, there can be one or two minima of $F$,
depending on temperature.
The crossover between classical and quantum regimes occures when
the two minima have the same free energy, which for $h_x=0.1$ takes place at 
$T_0^{(1)}=1.078 T_0^{(2)}$.

The crossover temperature for the escape rate is frequently
estimated by equating the ground-state tunneling exponent to 
that of thermal activation:
\begin{equation}\label{ground}
S(E_{\rm bottom})\equiv B = \frac{ U_{\rm max} - U_{\rm min} }{T_0^{(0)}}.
\end{equation}
The ground-state tunneling exponent $B$ given by Eq. (\ref{action})
can be analytically calculated \cite{enzsch86}, which
together with Eq. (\ref{uxbarr}) yields
\begin{eqnarray}\label{t00}
&&
T_0^{(0)} = \frac{SD}{4} \frac{ (1-h_x)^2 }
{ \ln\left( \displaystyle \frac{1+\sqrt{1-h_x^2}}{h_x}\right) - 
\sqrt{1-h_x^2} } 
\nonumber\\
&&
\renewcommand{\arraystretch}{2}
\qquad
{}\cong \frac{SD}{4}
\left\{
\begin{array}{ll}
\displaystyle
\frac{1}{\ln[2/(eh_x)]},         
                 & h_x\ll 1  \\
\displaystyle
\frac{3}{8^{1/2}}(1-h_x)^{1/2} ,
                 & 1-h_x\ll 1 ,    
\end{array}
\right. 
\end{eqnarray}
One can see from Fig. 1b that $T_0^{(0)}$ somewhat underestimates the
crossover temperature.
For $h_x=0.1$ one has $T_0^{(0)}=1.061 T_0^{(2)}<T_0^{(1)}$.
The estimation $T_0^{(0)}$ becomes, however, accurate in the
limit of small $h_x$.
The dependence of the crossover temperature $T_0$ on the
transverse field in the whole range, $0<h_x<1$, is presented in
Fig. 2. 
The temperature dependence of the escape rate can be
conveniently written in the form 
$\Gamma \sim \exp[-(U_{\rm max} - U_{\rm min})/T_{\rm eff}(T)]$,
where the dependence of the effective temperature on $T$ is presented 
in Fig. 3 for different $h_x$.
It can be seen from Fig. 3 that the most significant difference
between the crossover temperature $T_0^{(0)}$ of Eq. (\ref{ground}) and the
actual crossover temperature $T_0$ arises in the limit of small
barrier, that is, at $h_x\to 1$.
The former is described by the intersection of the dotted Arrhenius
line with the horizontal line corresponding to $T_{\rm eff}(T)/T_0$ at zero
temperature. 
From Eqs. (\ref{t02}) and (\ref{t00}) for $h_x\to 1$ one obtains 
$T_0^{(0)}/T_0^{(2)}=3\pi/(8\sqrt{2})\approx 0.833$.

\begin{figure}[t]
\unitlength1cm
\begin{picture}(11,7)
%
\centerline{\epsfig{file=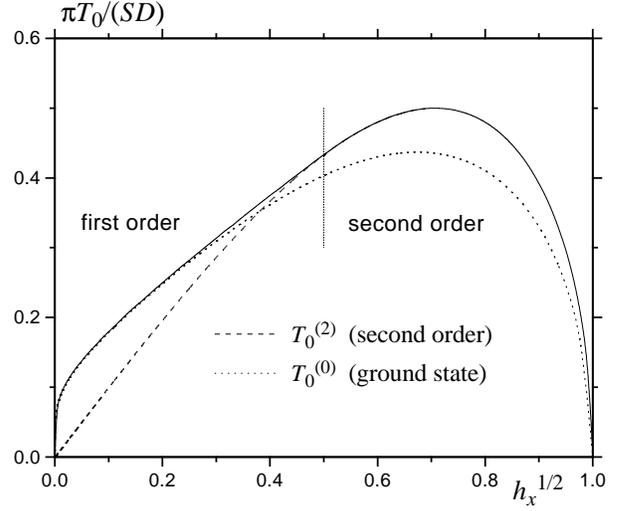,angle=-90,width=12cm}}
\end{picture}
\caption{ \label{tat_f2} 
Dependence of the crossover temperature $T_0$ on the
transverse field.
}
\end{figure}
\par

\begin{figure}[t]
\unitlength1cm
\begin{picture}(11,7)
%
\centerline{\epsfig{file=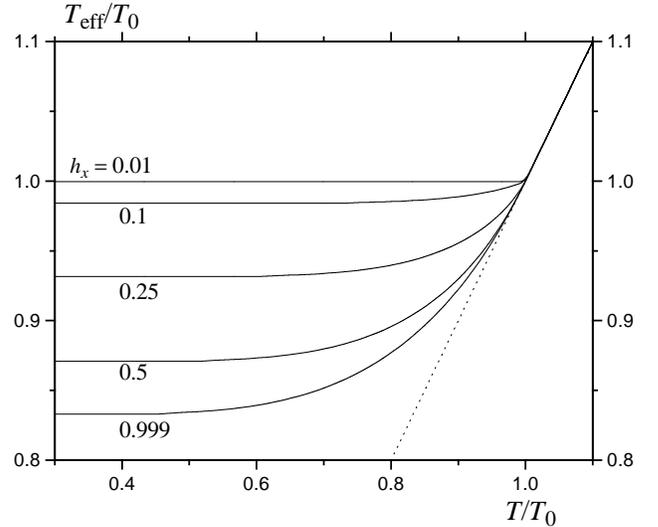,angle=-90,width=12cm}}
\end{picture}
\caption{ \label{tat_f3} 
Dependences of the effective temperature $T_{\rm eff}$ on $T$
for the different values of the transverse field.
}
\end{figure}

As follows from Fig. 3, the difference between the curves
$T_{\rm eff}(T)$ describing the first- and second-order
crossover is quite dramatic.
It must be easily observed in experiment if the appropriate
system is found.
Very recently experiments on individual small magnetic particles
with $S\sim 10^5-10^6$ have become possible \cite{weretal97q}.
In these experiments the barrier is lowered by tuning the
magnetic field to the critical value.
In our model this is the case of the second-order transition.
In order to get the first-order transition, $H_x$ must be lower than
$H_A/4$, where $H_A\equiv 2SD/(g\mu_B)$ is the anisotropy field.
This case requires a moderate spin $S$ in order to provide a
significant escape rate.
The Hamiltonian (\ref{sham}) has been found to be a good model
for Mn$_{12}$Ac \cite{sch97}, $S=10$.
In this case the quantization of spin levels becomes important.
However, our statement regarding the possibility of first- and
second-order transitions remains valid \cite{garchu97}.

The analogy with phase transitions in the temperature dependence
of the escape rate formally exists only in the limit of $S\to\infty$.
For a finite $S$, the transition from (\ref{gamint}) to (\ref{gammin})
has the accuracy of $1/S$.
Quantum corrections to the escape rate above $T_0$ \cite{haetalbor90} are
of the same order.
Thermal and quantum corrections will smoothen the first-order
transition in the narrow remperature region close to $T_0$.
Nevertheless, even for $S=10$, the difference between the
crossover at small and large $H_x$ must be easily observable.
The sharpness of the crossover between thermal and quantum
regimes also depends on the strength of the dissipation.
In the case of the low dissipation which is common for the
magnetic systems, its effect on the crossover is small \cite{haetalbor90}.

For Mn$_{12}$Ac the anisotropy field is about 10~T.
The crossover from thermal to quantum regime should, therefore,
switch from first to second order at $H_x\simeq 2.5$~T.
The crossover temperature is about 1~K \cite{garchu97}.
These ranges of field and temperature are easily accessible in experiment.
Note that similar effects may exist in the Fe$_8$ molecular magnet
where the crossover from thermal to quantum regime has been
already observed \cite{sanetal97}.
This system, however, is described by the spin Hamiltonian with
the transverse anisotropy which requires separate theoretical investigation.
We believe that the statement made in this paper, regarding the
possibility of first- and second-order crossover from thermal to
quantum regime, must be very general for spin systems.

This work has been supported by 
the U.S. National Science Foundation through
Grant No. DMR-9024250.

\renewcommand{\thefootnote}{\fnsymbol{footnote}}
\footnotetext[1]{ 
Electronic address: chudnov@lcvax.lehman.cuny.edu}

\renewcommand{\thefootnote}{\fnsymbol{footnote}}
\footnotetext[2]{ Permanent address: 
I. Institut f\"ur Theoretische Physik, Universit\"at Hamburg,
Jungiusstr. 9, D-20355 Hamburg, Germany.\\
Electronic address: garanin@physnet.uni-hamburg.de }


\begin{thebibliography}{10}

\bibitem{kra40}
{H. A. Kramers}, Physica {\bf 7},  284  (1940).

\bibitem{wen26}
{G. Wentzel}, Z. Phys. {\bf 38},  518  (1926).

\bibitem{kra26}
{H. A. Kramers}, Z. Phys. {\bf 39},  828  (1926).

\bibitem{bri26}
{L. Brillouin}, C. R. Acad. Sci. Paris {\bf 183},  24  (1926).

\bibitem{haetalbor90}
{P. H\"anggi, P. Talkner, and M. Borkovec}, Rev. Mod. Phys. {\bf 62},  251
  (1990).

\bibitem{gol59}
{V. I. Goldanskii}, Dokl. Acad. Nauk. USSR {\bf 124},  1261  (1959)
[Sov. Phys. Dokl. {\bf 4}, 74 (1959)].

\bibitem{aff81}
{I. Affleck}, Phys. Rev. Lett. {\bf 46},  388  (1981).

\bibitem{larovc83}
{A. I. Larkin and Yu. N. Ovchinnikov}, Pis'ma Zh. Eksp. Teor. Fiz. {\bf 37},
  322  (1983)
[Sov. Phys. JETP Lett., {\bf 37}, 382 (1983)].  

\bibitem{chu92}
{E. M. Chudnovsky}, Phys. Rev. A {\bf 46},  8011  (1992).

\bibitem{enzsch86}
{M. Enz and R. Schilling}, J. Phys. C {\bf 19},  L711  (1986).

\bibitem{lvhsuto86}
{J. L. van Hemmen and A. S\"ut\H o}, Physica B {\bf 141},  37  (1986).

\bibitem{schwrelvh87}
{G. Scharf, W. F. Wreszinski, and J. L. van Hemmen}, J. Phys. A {\bf 20},  4309
   (1987).

\bibitem{chugun88}
{E. M. Chudnovsky and L. Gunter}, Phys. Rev. Lett. {\bf 60},  661  (1988).

\bibitem{zas90plazas90prb}
{O. B. Zaslavskii}, Phys. Lett. A {\bf 145},  471  (1990);
 Phys. Rev. B {\bf 42},  992  (1990).  

\bibitem{gar91jpa}
{D. A. Garanin}, J. Phys. A {\bf 24},  L61  (1991).

\bibitem{sch97}
{M. A. Novak and R. Sessoli},  in {\em Quantum Tunneling of Magnetization},
  edited by L. Gunther and B. Barbara (Kluwer, Dordrecht, 1995);
{B. Barbara {\em et al}}, J. Magn. Magn. Mater. {\bf 140-144}, 1825 (1995);
{J. R. Friedman, M. P. Sarachik, J. Tejada, and R. Ziolo}, Phys. Rev. Lett.
  {\bf 76},  3830  (1996); see also  
{B. Schwarzschild}, Physics Today {\bf 50}, No. 1,  17  (1997).

\bibitem{scharf74}
{G. Scharf}, Ann. Phys. (N. Y.) {\bf 83}, 71 (1974).

\bibitem{zasulytsu83}
{O. B. Zaslavskii, V. V. Ulyanov, and V. M. Tsukernik}, Fiz. Nizk. Temp. 
{\bf 9},  511  (1983)
[Sov. J. Low Temp. Phys. {\bf 9}, 259 (1983)].

\bibitem{fey72}
{R. P. Feynman}, {\em Statistical Mechanics} (Benjamin, New York, 1972).

\bibitem{calleg83}
{A. O. Caldeira and A. J. Leggett}, Ann. Phys. (N.Y.) {\bf 149},  374  (1983).

\bibitem{grawei84}
{H. Grabert and U. Weiss}, Phys. Rev. Lett. {\bf 53},  1787  (1984).

\bibitem{larovc84}
{A. I. Larkin and Yu. N. Ovchinnikov}, Zh. Eksp. Teor. Fiz. {\bf 86},  719
  (1984)
[Sov. Phys. JETP, {\bf 59}, 420 (1984)].

\bibitem{zwe85}
{W. Zwerger}, Phys. Rev. A {\bf 31},  1745  (1985).

\bibitem{rishaefre85}
{P. S. Riseborough, P. H\"anggi, and E. Freidkin}, Phys. Rev. A {\bf 32},  489
  (1985).

\bibitem{weretal97q}
{W. Wernsdorfer, K. Hasselbach, E. Bonet Orozco, A. Benoit, D. Mailly, O. Kubo,
  and B. Barbara}, Phys. Rev. Lett.  {\bf 79} 4014 (1977).

\bibitem{garchu97}
{D. A. Garanin and E. M. Chudnovsky}, Phys. Rev. B 
{\bf 56} 11102 (1997).

\bibitem{sanetal97}
{C. Sangregorio, T. Ohm, C. Paulsen, R. Sessoli, and D. Gatteschi}, Phys. Rev.
  Lett. {\bf 78},  4645  (1997).

\end{thebibliography}

\end{document}